\documentclass[a4paper,aps,prd,groupedaddress,preprintnumbers,showpacs,10pt]{revtex4}%nofootinbib,
\usepackage{graphicx}
\usepackage{latexsym}
\usepackage{amsfonts}
\usepackage{amssymb}
\usepackage{amsmath}
\usepackage{slashed}
\usepackage{tabularx,url,color}
\usepackage{hyperref}
\usepackage{multirow}
\usepackage{longtable,lscape}
\usepackage{graphicx,booktabs}
\usepackage{epsfig}

\newcommand{\mev}{{\ensuremath \text{MeV}}}
\newcommand{\gev}{{\ensuremath \text{GeV}}}

\begin{document}
\title{Study on the structures of the four-quark states in terms of the Born-Oppenheimer approximation}

\author{Xue-Wen Liu$^{1}$\footnote{liuxuewen@mail.nankai.edu.cn}, Hong-Wei Ke$^{2}$\footnote{khw020056@hotmail.com}, Yi-Bing Ding$^{3}$\footnote{ybding@gucas.ac.cn} and Xue-Qian Li$^{1}$\footnote{lixq@nankai.edu.cn}}

\affiliation{  $^{1}$ School of Physics, Nankai University, Tianjin 300071, China \\
  $^{2}$ School of Science, Tianjin University, Tianjin 300072, China\\
  $^{3}$ School of Physics, University of Chinese Academy of Sciences, Beijing 100049, China}

%\date{\today}

\begin{abstract}
In this work, we use the Born-Oppenheimer approximation where the potential between atoms can be approximated as a function
of distance between the two nuclei  to study the four-quark bound states. By the approximation, Heitler and London calculated
the spectrum of hydrogen molecule which includes two protons (heavy) and two electrons (light). Generally, the observed exotic mesons
$Z_b(10610)$, $Z_b(10650)$, $Z_c(3900)$ and $Z_c(4020)$($Z_c(4025)$) may be
molecular states made of two physical mesons and/or in diquark-anti-diquark structures. In analog to the Heitler-London method for calculating the mass of
hydrogen molecule, we investigate whether there exist energy minima for these two structures.
By contrary to the hydrogen molecule case where
only the spin-triplet possesses an energy minimum, there exist minima for both of them. It implies that
both molecule and tetraquark states can be stable objects. But since they have the same quantum numbers, the two states may mix to result
in the physical states. A consequence would be that partner exotic states co-existing with $Z_b(10610)$,
$Z_b(10650)$, $Z_c(3900)$ and $Z_c(4020)$($Z_c(4025)$) are predicted and should be experimentally observed.

\end{abstract}

\maketitle

%%%%%%%%%%%%%%%%%%%%%%%%%%%%%%%%%%%%%%%%%%%%%%%%%%%%%%%%%%%%%%%%%%%%%%%%%%%%%%%
\section{Introduction}
\label{sec:intro}
The naive quark model suggests that a meson is
made of a quark and an anti-quark whereas a baryon consists of three
quarks. The constituents in the hadrons are bound together by the
QCD interaction to constitute a color singlet. But neither the
quark model nor the QCD theory ever forbids existence of
multi-quark states as long as they are color-singlets. The fact
that by several years of hard work all experimental trials  to
observe pentaquarks failed, greatly discouraged theorists and
experimentalists of high energy physics even though the idea about
pentaquarks is really stimulating. One may ask if the nature
indeed only favors economic textures of hadrons. The situation
changes by the discovery of the exotic states
$Z_b(10610)$ and $Z_b(10650)$ \cite{Collaboration:2011gja}, and especially the newly
observed $Z_c(3900)$ \cite{Ablikim:2013mio},
$Z_c(4020)$ \cite{Ablikim:2013wzq} and
$Z_c(4025)$ \cite{Ablikim:2013emm}. The characteristics of such
states are that $Z_b$ and $Z_c$-mesons contain hidden bottom $b\bar b$ or charm
$c\bar c$ respectively and both are charged,
therefore they cannot be simply $b\bar b$ or $c\bar c$ bound
states, but multi-quark states and are called as exotic states compared to
the regular structures.

The inner structure of the multi-quark states is more complicated than the
regular mesons that
the exotic states can be molecular states
or tetraquarks or their mixtures. The molecular state is constructed by two color
singlet mesons. A strong point to support such a structure is that
the mass of the newly discovered meson $Z_b(10610)$ is close to
the sum of the masses of $B$ and $\bar{B}^*$, while the mass of
$Z_c(3900)$ is also close to a sum of $D$ and $D^*$ masses.
Instead, the study on the decay modes of such mesons seem to
support the tetraquark structure \cite{Ke1,Ke2}. To clarify the
structures of those exotic states, one may need to
investigate their all characteristics based on fundamental dynamics, instead of simply considering closeness of
their masses with the sum of the constituents.

One observation may call our attention. The resonances $Z_b(10610)$,
$Z_b(10650)$, $Z_c(3900)$, $Z_c(4020)$ and $Z_c(4025)$ have been experimentally observed and
confirmed as exotic four-quark states. Many authors \cite{MoZb1,MoZb2,MoZb3,MoZc1,MoZc2} assumed them to be molecular states of
$B$, $B^*$, $D$ and $D^*$ (and the corresponding anti-mesons) which are well experimentally measured.
There is a common point that the masses of the observed exotic mesons are larger than the sum of the supposed constituent mesons.
Concretely, $10608.4\pm 2.0$ MeV \cite{Collaboration:2011gja} (the mass of $Z_b(10610)$) is larger than sum of masses of $B$ and $B^*$ (10604.45 MeV);
$10653.2\pm 1.5$ MeV \cite{Collaboration:2011gja} (the mass of $Z_b(10650)$) is larger than sum of masses of $B^*$ and $B^*$ (10650.4 MeV);
$3899\pm 3.6\pm 4.9$ MeV  \cite{Ablikim:2013mio} (the mass of $Z_c(3900)$) is larger than sum of $D$ and $D^*$ (3876.6 MeV);
$4022.9\pm 0.8\pm 2.7$ MeV \cite{Ablikim:2013wzq} (the mass of $Z_c(4020)$) is larger than sum of $D^*$ and $D^*$ (4013.96 MeV). Generally, unless there exists
a linearly increasing potential (such as the confinement potential for quarks) or a barrier, the binding energy of two
constituent mesons which is caused by exchanging color-singlet hadrons must be negative. Thus, by the common sense
the mass of a composite meson should be smaller than the sum of the two (or more) constituent masses. Moreover as estimated by some authors \cite{Maiani:2004vq,Kleiv:2013dta}
the masses of those exotic states are also larger than the sums of masses of the concerned diquark and anti-diquark.
In our calculation, even though the sum of the two diquark masses is larger than the mass of the corresponding exotic meson, the negative binding energy
still makes the resultant total energy smaller than the exotic meson. That may imply neither
molecular nor teraquark states alone correspond to the observed exotic mesons. Our study indicates that only their mixture
provides a reasonable picture for the four-quark states.
Thus, both molecular state and teraquark state should exist, even though they may not be physical states which
we observe in experiments.

By the Born-Oppenheimer \cite{BO} approximation, the potential
between atoms can be approximated as a function of the distance
between two nuclei, by the scheme, Heitler and London \cite{HL} calculated the spectrum of
hydrogen molecule. In that case, the two protons are supposed to
be at rest and the two electrons are moving. Since the two electrons are identical fermions,
the wavefunction of the two-electron system  must be
totally anti-symmetric.
%Their spin-component can be either symmetric or
%anti-symmetric, i.e. the two electrons are in spin triplet or
%singlet. Accordingly by the exchange symmetry the spatial
%wavefunction must also be anti-symmetric or symmetric. Taking those
%two anti-symmetrized wavefunctions as trial functions the
%expectation value of the total Hamiltonian as a
%function of the distance between the two protons is calculated. In this
%approximation, the interaction between an electron and a proton which belong to different hydrogen atoms and that between two electrons are
%treated as a perturbation.
It was found that there is only one energy
minimum corresponding to the triplet. Namely, in the hydrogen
molecule the two electrons must be in the spin-triplet.

Comparing with the hydrogen molecule, $Z_b$ (or $Z_c$) is made of four quarks: $Q,\;\bar Q,\; u(\bar u),\; \bar d(d)$, where $Q$ stands for $b$ or $c$ quark, since $Q,\;\bar Q$ are much heavier than
the light flavors, so we can approximate them to be at rest. Thus it is natural that we can separate the four quarks into two groups. One
possibility is that each group is in a color singlet, which corresponds to a molecular state, whereas another possibility is that one group
containing $Q u$ is in a color-anti-triplet(or a sextet)  and the other group containing $\bar Q\bar d$ is in a color-triplet (or an anti-sextet), i.e. the dipole-anti-dipole structure.
Since $u$ and $\bar d$ are not identical particles,
the wavefunction does not need to be anti-symmetrized.
%The two trial wave functions in this case are products of
%two mesons and the dipole-anti-dipole.
By the Born-Oppenheimer approximations, the potential between two groups can be a function
of distance between $Q$ and $\bar Q$ and interactions between the two groups are taken as a perturbation. Since the interactions between quarks are complicated,
calculation of the energy spectrum of the exotic state is much more difficult than for hydrogen molecule.
It is noted that Braaten et al.\cite{Braaten:2014qka} also consider the Born-Oppenheimer potential to deal with the four-quark states.

First we need to determine the wavefunctions of the color singlet
of $Q\bar d (\bar Q u)$ and the color-anti-triplet (or sextet) dipole $Qu$(color-triplet or anti-sextet $\bar
Q\bar d)$. Here we use the  Cornell potential \cite{Cornell} as the
interaction between the quarks and since the light flavors are
relativistic, following the literature, we employ the Schr\"odinger-like equation with relativistic kinematics.
The effective interaction between the quarks (quark-anti-quark) which belong to different groups
is complicated, because not only the
short-distance QCD interaction exists, but also the long-distance
interaction which can be treated by exchanging color-singlet light
mesons: $\pi,\; \sigma$ and $\rho$ etc. (for the molecule case) or the color-flux tube (for the tetraquark case) plays important roles. Here we do not involve the
strange flavor. In analog to the hydrogen molecule, we calculate
the spectrum of the ground state of the four-quark system (the molecule and tetraquark separately). Our strategy
is similar to the Heitler-London approximation,
namely we take the products
of the two meson wavefunctions(for the molecule) and diquark-anti-diquark wavefunctions (for the tetraquark) as two
independent trial functions and calculate the interaction between
two groups to obtain the spectra as functions of the distance
between $b$ and $\bar b$ ($c$ and $\bar c$). Our goal is to see whether the molecular
state or tetraquark state can possess energy minima with
respect to the distance between $Q$ and $\bar Q$, by which one can
judge if molecule or tetraquark can be physically allowed. If there
exist minima for both cases, we would conclude that both
structures are probable and the real physical state could be a
mixture of the two structures. (In fact, our computations confirm
that there are minima for both.)

This work is organized as follows, after this long introduction, we would formulate the expressions for the energy spectra. We first present  relevant
effective potentials for the meson and diquark which are composed of a heavy quark and a light flavor  and then
derive the Born-Oppenheimer potentials for both molecule and tetraquark.
In Sec. \ref{formulae}, we discuss the explicit color, spin structures of the molecular and tetraquark states and solving the Schr\"odinger-like equation to obtain
the spatial wavefunctions of color-singlet meson and color-anti-triplet diquark.
%Then we treat the interaction between the constituents residing in different groups by
%perturbation, finally, the Born-Oppenheimer potential with respect to distance and get the minima.
In Sec. \ref{results}, along with all input parameters we present our numerical results which show that for both
molecule and tetraquark there exist minima with repect to the distance between $Q$ and $\bar Q$. The last section is
devoted to our discussion and conclusion.

%%%%%%%%%%%%%%%%%%%%%%%%%%%%%%%%%%%%%%%%%%%%%%%%%%%%%%%%%%%%%%%%%%%%%%%%%%%%%%%
\section{Derivation of the concerned Formulae}
\label{formulae}
In this section, we derive the theoretical formulae for calculating the mass spectra and wavefunctions of both molecular and tetraquark states. We first by solving the Schr\"odinger-like equations to
obtain the mass spectra and wavefunctions of the mesons $B,\;B^*,\;D$ and $D^*$ and diquark (anti-diquark) which would be the trial functions for later calculation. But it is noted, since diquark is not a physical state, we determine its mass spectrum and wavefunction
via theoretical computations. We go on using the Born-Oppenheimer approximation to evaluate the mass spectra of molecular
and tetraquark states which are functions of the distance between two heavy constituents $Q$ and $\bar Q$.

\subsection{The potentials in various cases }
\label{potentials}
\begin{figure}[!htbp]
\begin{centering}
\includegraphics[width=0.75\textwidth]{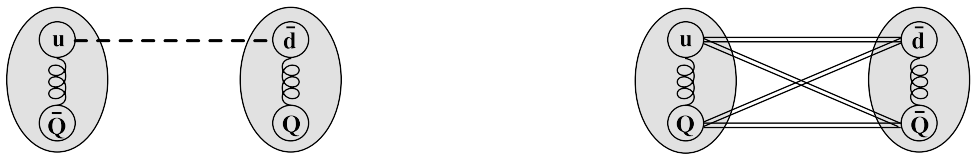}
\caption{The configurations of four-quark system(the left panel is the molecular state and the right one is the tetraquark state)}
\label{fig:config}
\end{centering}
\end{figure}

Here we first obtain the effective potentials between concerned constituents inside a color-singlet, i.e. mesons and color-triplet (anti-triplet), i.e. anti-diquark(diquark).
Then we go on to derive the potential between constituents coming from different groups. For the
two distinct configurations(the molecular state and tetraquark (diquark-antidiquark) state (see Fig. \ref{fig:config}), the effective interactions are
different.

\subsubsection{For meson and diquark(antidiquark) states}
In this subsection, let us first discuss the interactions among the constituents inside a meson ($q\bar Q(\bar q Q)$) or a (anti)diquark($q Q(\bar q \bar Q)$).
The general Hamiltonian
can be written as
\begin{equation}\label{eq:Ham}
 H=\sqrt{{\bf p}^2_i+m_i^2}+\sqrt{{\bf P}_j^2+m_j^2}+V(r) ,~~~~~i=q(\bar q); j=Q(\bar Q)\\
\end{equation}
where the ${\bf p}_i$ and ${\bf P}_j$ are the 3-momenta of the light flavor $q(\bar q)$ and heavy one $Q(\bar Q)$ respectively. The interaction potential is
\begin{equation}\label{eq:V}
V(r_{ij})=V_{\text{oge}}(r_{ij})+V_{\text{con}}(r_{ij}),
\end{equation}
and $r_{ij}$ is the distance between quarks (quark-antiquark). The one-gluon exchange(oge) term $V^{oge}(r_{ij})$, which plays the main role at short distance,  is \cite{Fan-wang}
\begin{equation}\label{eq:Voge}
V_{\text{oge}}(r_{ij})=\frac{1}{4}\alpha_s({\mbox{\boldmath $ \lambda$}}_i^c\cdot{\mbox{\boldmath $ \lambda$}}_j^c)
\Big[\frac{1}{r_{ij}}-\frac{\pi}{2}\Big(\frac{1}{m_i^2}+\frac{1}{m_j^2}+\frac{4}{3m_im_j }{\mbox{\boldmath $\sigma$}}_i\cdot{\mbox{\boldmath$\sigma$}}_j\Big) \delta (r_{ij})\Big],
\\
\end{equation}
and the confinement piece $V_{\text{con}}(r_{ij})$ takes the linear form\cite{Cornell} as
\begin{equation}\label{eq:Vcon}
V_{\text{con}}(r_{ij})=-\frac{1}{4}({\mbox{\boldmath$\lambda$}}_i^c\cdot{\mbox{\boldmath$\lambda$}}_j^c)
    (b r_{ij}+c)
\end{equation}
where ${\mbox{\boldmath$\lambda$}}_i^c$ and ${\mbox{\boldmath$\sigma$}}_i$ are respectively the color $SU_c(3)$ and spin operators acting on quark $i$,
and $m_i$ is the quark mass.  $b$ is the string tension,  and $c$ is a global zero-point energy.  $\alpha_s$ is the QCD running coupling constant,
which depends on the re-normalization scale $\mu^2$ \cite{Vijande:2004he}
\begin{equation}\label{running}
\alpha_s(\mu^2)=\frac{\alpha_0}{\ln(\frac{\mu^2+\mu^2_0}{\Lambda_0^2})},
\end{equation}
where $\mu=m_im_j/(m_i+m_j)$ is the reduced mass of the $q_i\bar{Q}_j$ system and  $\Lambda_0$, $\alpha_0$, $\mu_0$  are fitted parameters. The framework
can be generalized to the case for diquark(anti-diqaurk) which involves two quarks (two anti-quarks).

The $\delta$-function in Eq.(\ref{eq:Voge}) is replaced by a Gaussian smearing function \cite{Weinstein:1983gd} with a fitted parameter $h$
\begin{equation}\label{eq:delta}
\delta(r_{ij})\rightarrow \frac{h^3}{\pi^{3/2}}e^{-h^2r_{ij}^2}. \\
\end{equation}
\subsubsection{For molecular states}

Now we specify the interaction between the two mesons for the molecular structures(see the left panel of Fig. \ref{fig:config}). Since the constituent mesons are in color singlet, the quarks (antiquarks) in one meson do not interact with the quarks in another meson via exchanging a single gluon, thus the interaction between $B^{(*)}B^{(*)}$ (or $D^{(*)}D^{(*)}$) only comes from the meson-exchange between the light flavors $q\bar{q}$.

The constituent quark model was thoroughly studied by many authors, for example, Vijande et al. \cite{Vijande:2003ki,Vijande:2004he}, and its successful applications to phenomenology are noted, thus here we employ it to derive the effective interaction between mesons.
The interactions $V_{\text{me}}(r_{ij})$ induced by meson-exchange(me) between $q$ and $\bar q$ includes contributions of pseudoscalar$(p)$ and scalar$(s)$,
\begin{align}\label{eq:me}
  V_{\text{me}}(r_{ij}) &
    =\sum_{a=1}^3V_\pi(r_{ij}){\bf F}_i^a\cdot {\bf F}_j^a  +\sum_{a=4}^7V_K(r_{ij}){\bf F}_i^a\cdot {\bf F}_j^a \nonumber \\
   & +V_\eta(r_{ij})[\cos\theta_p({\bf F}_i^8\cdot {\bf F}_j^8)-\sin\theta_p]
    +V_{\sigma}(r_{ij}),
\end{align}
the explicit form of the  interactions are
\begin{equation}\label{eq:vps}
  V_{\chi}(r_{ij})=\frac{g^2_{ch}}{4\pi}\frac{m_\chi^2}{12m_im_j}\frac{\Lambda^2_\chi}
  {\Lambda^2_\chi-m^2_\chi}m_\chi\Big[Y(m_\chi r_{ij})-\frac{\Lambda^3_\chi}{m^3_\chi}Y(\Lambda_\chi r_{ij})\Big]
  ({\mbox{\boldmath$\sigma$}}_i\cdot{\mbox{\boldmath$\sigma_j$}}),
\end{equation}

\begin{equation}\label{eq:vs}
  V_{\sigma}(r_{ij})=-\frac{g^2_{ch}}{4\pi}\frac{\Lambda^2_\sigma}
  {\Lambda^2_\sigma-m^2_\sigma}m_\sigma\Big[Y(m_\sigma r_{ij})-\frac{\Lambda_\sigma}{m_\sigma}Y(\Lambda_\sigma r_{ij})\Big]
\end{equation}
with $\chi=\pi, K, \eta$ and ${\bf F}_{i}^{a} (a=1,2,\cdots,8)$ are the $SU(3)$ flavor matrices. $Y(x)=e^{-x}/x$ is the Yukawa function, $g_{ch}$ is the chiral coupling constant, $\theta_p$ is the mixing angle for the physical $\eta$ and $\eta'$, and $\Lambda 's $ are the chiral symmetry breaking scales.
Once the potential between $qq$ is determined, the corresponding potential for  $q\bar{q}$ can also be obtained from a $G$-parity transformation \cite{zhangzx}.
It is noted that the employed framework is the SU(3) chiral quark model where the  heavy quark ($c$ or $b$) does not couple to the SU(3) mesons.

Furthermore, in the effective potential there also exists a piece $V_{\text{ann}}(r_{ij})$ induced by quark-antiquark $(q,\bar{q})$ pair annihilation into a light mesons which mediate
interactions in s-channel. To the lowest order the quark-antiquark
pair resides in an S-wave state, the contribution of $\sigma$-meson($J^{pc}=0^{++}$) can be neglected \cite{zhangzx}. So here we only keep the contributions of $\pi$  and  $\rho$
to the potential \cite{Faessler82,Entem-ann}
\begin{equation}\label{eq:vApsR}
V_{\text{ann},\pi}(r_{ij})=\frac{g_{ch}^2\delta(r_{ij})}{4m_q^2-m_{\pi}^2}
\Big(\frac{1}{3}+\frac{1}{2}{\mbox{\boldmath$\lambda$}}^c_q\cdot{\mbox{\boldmath$\lambda$}}
^c_{\bar{q}}\Big)
\Big(\frac{1}{2}-\frac{1}{2}{\mbox{\boldmath$\sigma$}}_q\cdot{\mbox{\boldmath $\sigma$}}_{\bar{q}}\Big)
\Big(\frac{3}{2}+\frac{1}{2}{\mbox{\boldmath$\tau$}}_q\cdot{\mbox{\boldmath$\tau$}}_{\bar{q}}\Big),\\
\end{equation}
and
\begin{equation}\label{eq:vAvR}
V_{\text{ann},\rho}(r_{ij})=-\frac{g_{v}^2\delta(r_{ij})}{4m_q^2-m_{\rho}^2}
\Big(\frac{1}{3}+\frac{1}{2}{\mbox{\boldmath$\lambda$}}^c_q\cdot{\mbox{\boldmath$\lambda$}}
^c_{\bar{q}}\Big)
\Big(\frac{3}{2}+\frac{1}{2}{\mbox{\boldmath$\sigma$}}_q\cdot{\mbox{\boldmath $\sigma$}}_{\bar{q}}\Big)
\Big(\frac{3}{2}+\frac{1}{2}{\mbox{\boldmath$\tau$}}_q\cdot{\mbox{\boldmath$\tau$}}_{\bar{q}}\Big).\\
\end{equation}
where ${\mbox{\boldmath$\tau$}}$ is the isospin operator, and the $\delta$-function is also rewritten in the same form as Eq.\ref{eq:delta}.

Summing all the individual pieces up, the interaction between the two mesons of the molecule is
\begin{equation}\label{eq:moH}
 H^{\text{(mol)}}_{\text{int}}= V_{\text{me}}(r_{ij})+V_{\text{ann}}(r_{ij}).
\end{equation}

\subsubsection{For tetraquark states}

Now, for the case of tetraquark, we are dealing with the interaction between the two groups $qQ$ and $\bar q \bar Q$. The key point
is to derive an effective potential. The total Hamiltonian is written as
\begin{equation}\label{eq:flux-con}
H^{\text{(tetra)}}_{\text{int}}=\sum_{i=u,b;\atop j=\bar d, \bar b} \Big[V_{\text{oge}}(r_{ij})+V'_{\text{con}}(r_{ij})\Big]
\end{equation}

The interaction among constituents in diquark and that in anti-diquark is not simply determined by the perturbative QCD, becuse the short-distance and
long-distance  contributions exist simultaneously. Following Brodsky et al.\cite{Brodsky:2014xia}, the flux tube model may properly describe the
interaction for the tetraquark case. Meanwhile in this case the contributions of the meson exchanges can be safely ignored comparing with that of
gluon exchange \cite{Carlson:1982xi}. The general form of Hamiltonian in the flux-tube model can also be decomposed into the Coulomb-type piece which is
responsible for short distance interaction and the confinement piece for the long-distance interaction.
As Brodsky et al.\cite{Brodsky:2014xia} suggested, in a ``substantial separation", diquark and antidiquark are connected by the flux-tube.
It is noted that in our pictures according to the Heitler-London approximation, we need to consider all the interactions among the constituents of different groups, thus
we account for the interactions as shown on the right panel of Fig. \ref{fig:config}.  Obviously, as summing over all the contributions a resultant Born-Oppenheimer potential
would be obtained which is also an effective flux tube between the diqaurk and anti-diquark and it is the picture of Ref.\cite{Brodsky:2014xia}.
Moreover, as is well known, when the tension on the
sting is beyond a certain bound the string would be broken into two stings and at the new ends a quark-anti-quark pair is created\cite{Kokoski:1985is,Kumano:1988ga}.  One can use
a step function to describe the breaking effect as
\begin{equation}
 (b r_{ij}+c)\theta(r-r_0),
\end{equation}
where $r_0$ is a parameter corresponds to the strengthening limit of the string. A typical scale for non-perturbative QCD is $\Lambda_{QCD}$, therefore
it is natural to consider $r_0=1/\Lambda_{QCD}$. Just as smearing the delta function, we need also to smear the step function. In fact
$$\theta(r-r_0)=\lim_{\varepsilon\to 0}\frac{1}{\text{e}^{\frac{1}{\varepsilon}(r_{ij}-r_0)}+1},$$
smearing the step function implies that we keep $\varepsilon$ as a non-zero free parameter to be determined.

Here the interaction between  $q$ of diquark and $\bar q$ from the antidiquark at a relatively large distance is described by a modified form as
\begin{equation}\label{eq:flux-con}
V'_{\text{con}}(r_{ij})=-\frac{1}{4}({\mbox{\boldmath$\lambda$}}_i^c\cdot{\mbox{\boldmath$\lambda$}}_j^c)
(b r_{ij}+c)\frac{1}{\text{e}^{\frac{1}{\varepsilon}(r_{ij}-r_0)}+1}
\end{equation}
and $\varepsilon$ is a parameter in fm, and we set $\Lambda_{QCD}=280\mev$ in this work.

%%%%%%%%%%%%%%%%%%%%%%%%%%%%%%%%%%%%%%%%%%%%%%%%%%%%%%%%%%%%%%%%%%%%%%%%%%%%%%%

\subsection{Wave functions of four-quark states  }
\label{sec:Wavefunction}
Combining all the degrees of freedom of the constituent quarks, the total wave function is a direct product of the radial, spin, color, and isospin(flavor) parts
\begin{equation}
\label{eq:wave}
\left|\psi_{\alpha}\right>=\left|
C_{\alpha}\right>\otimes\left|
I_{\alpha}\right>\otimes[\left|\phi_{\alpha}\right>
\otimes\left|S_{\alpha}\right>]^{JM},~~~~~\alpha=\text{(mol)},(\text{tetra}),
\end{equation}
for molecular state and tetraquark state separately.
Unlike the hydrogen molecules, the involved quarks (antiquarks) are not identical, so the Pauli principle does not impose any restrictions on the compositions.

\subsubsection{Radial wave function}
In the essence of the Born-Oppenheimer approximation, we can choose the product of the two clusters's wavefunctions as the basis shown in Fig. \ref{fig:config}
\begin{equation}\label{eq:R}
\phi_{\text{(mol)}}=\phi_{u\bar{Q}}\otimes\phi_{\bar{d}Q},~~ \phi_{\text{(tetra)}}=\phi_{uQ}\otimes\phi_{\bar{d}\bar{Q}}.
\end{equation}

The radial wave function for each cluster is obtained by solving the Schr\"odinger-like equation
\begin{equation}\label{eq:wave}
\big[\sqrt{{\bf p}_q^2+m^2_q}+\sqrt{{\bf P}_Q^2+m^2_Q}+V(r)\big]\phi_{\kappa}=E\phi_\kappa, ~~\kappa=u\bar{Q},\bar{d}Q,uQ,\bar{d}\bar{Q},
\end{equation}
where the potential $V(r)$ takes the Cornell type potential (see Eq.(\ref{eq:Voge}) and Eq.(\ref{eq:Vcon})), $m_{q}$ and $m_{Q}$ are the masses of light$(u, d)$ and heavy $c,b$ quarks.
It applies to both meson and diquark cases with different color factors.

\begin{figure}[!htbp]
\begin{center}
\begin{tabular}{ccc}
\scalebox{0.3}{\includegraphics{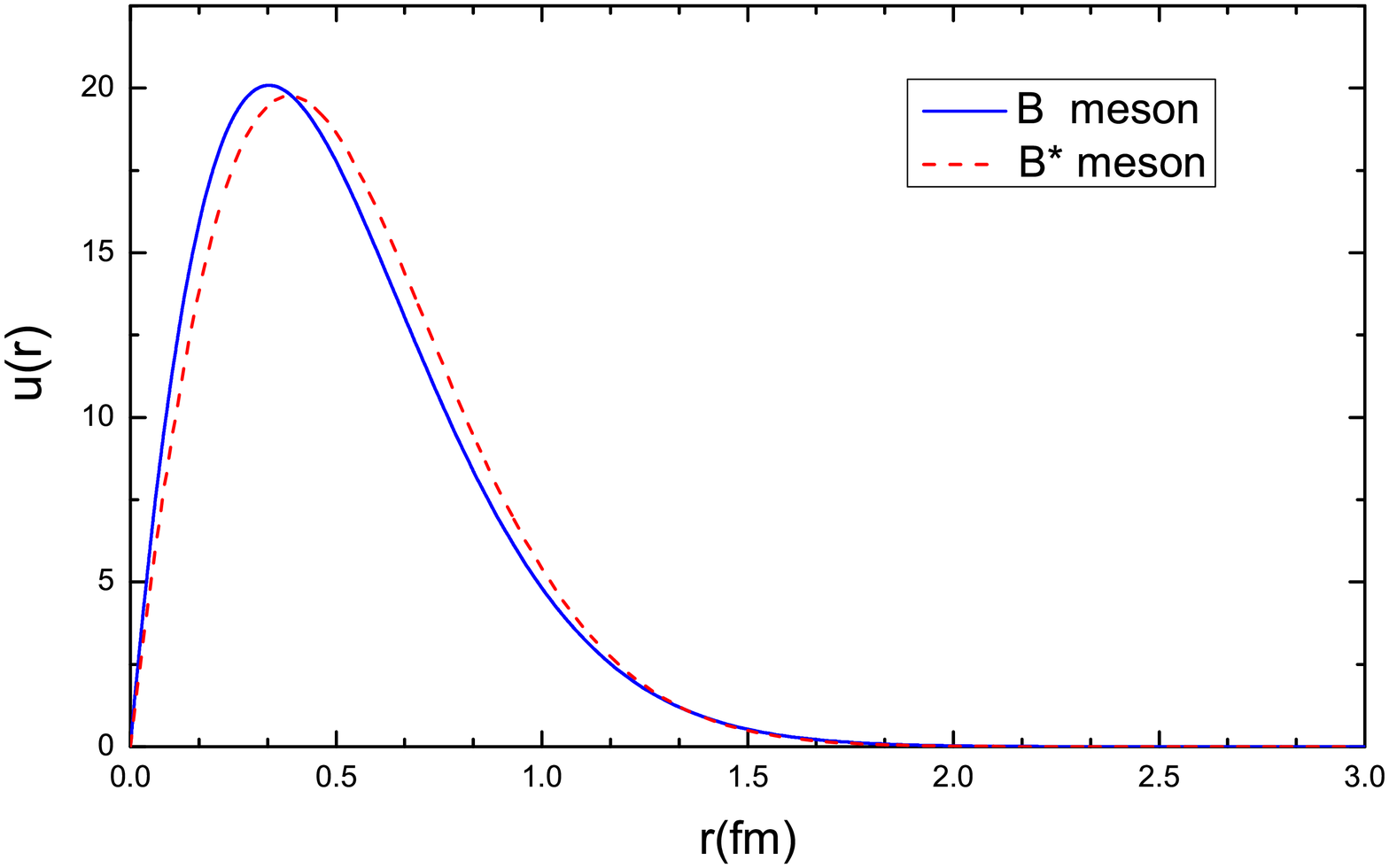}}\scalebox{0.3}{\includegraphics{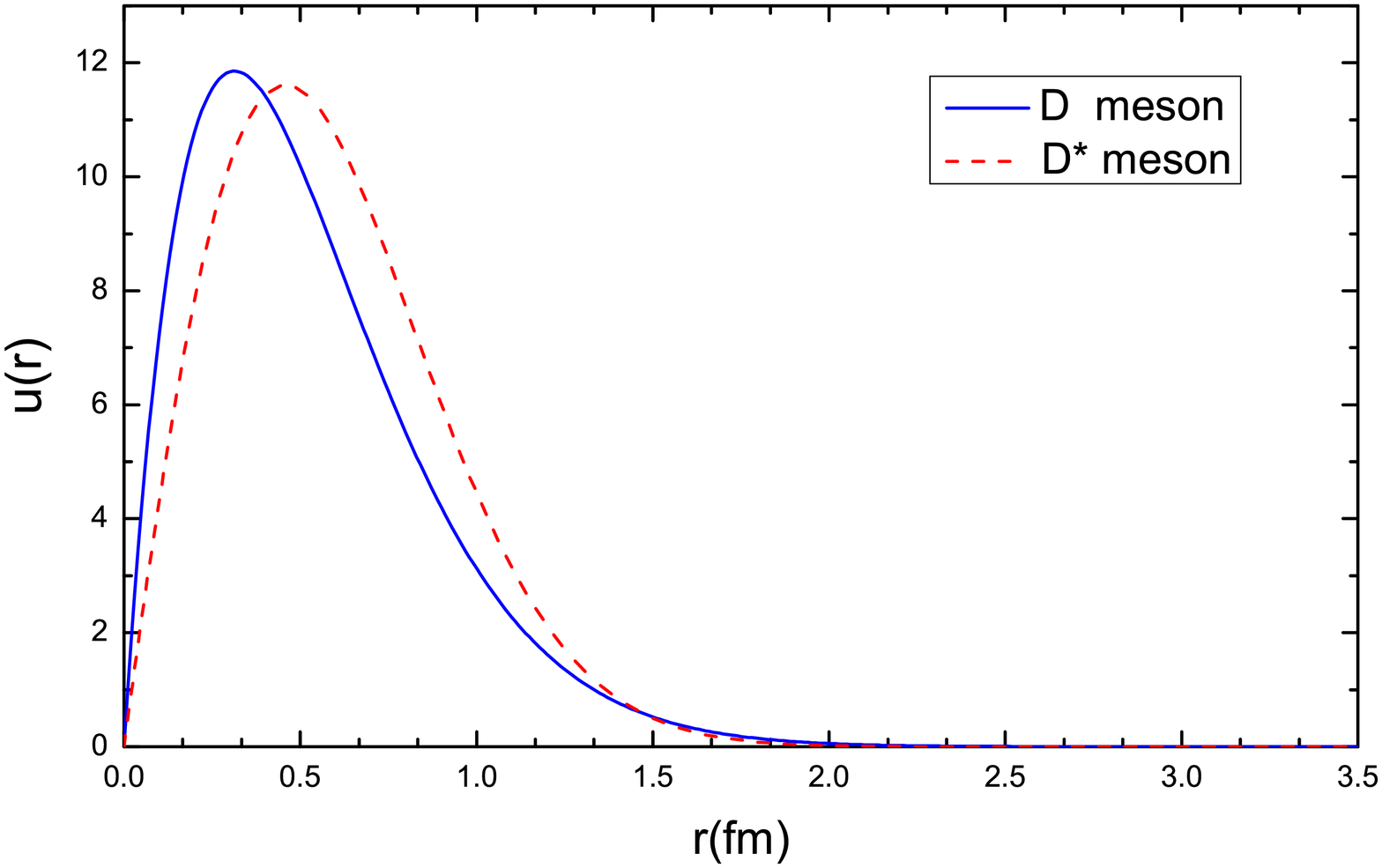}}
\\
\qquad(a)\qquad\qquad\qquad\qquad\qquad\qquad\qquad\qquad\qquad\qquad
\,\,\,\,\,\,\,\,\,\,\,\,\,\,\,\,\,\,\,\,\,\,\,(b)
\end{tabular}
\end{center}
\caption{The reduced wavefunction $u(r)$ in the coordinate space. (a) the solid curve is for  $B$ and the dashed one is for $B^*$ .(b)  the solid curve is for  $D$  and the dashed one is for  $D^*$  }\label{wave}
\end{figure}

We solve the  Schr\"odinger-like equation numerically in terms of the program offered by the authors of Ref. \cite{yang} to  deduce the radial wavefunction $u(r)$, defined as $\phi_{\kappa}({\bf r})=\frac{u_l(r)}{r} Y_{lm}(\hat {\bf r})$, with $l=0$. In Fig. \ref{wave} the wavefunctions of $B^{(*)}$ and $D^{(*)}$ are shown. The eigenvalues are given in Table. \ref{tab:mesons} where the constituent quarks masses are input parameters.

\begin{table}[!htbp]
\renewcommand{\arraystretch}{1.5}
\caption{Masses of heavy mesons and diquark(with spin-0 and spin-1) calculated by solving the  Schr\"odinger-like equation, and for a comparison experimental data\cite{Agashe:2014kda} and results from QCD sum rules,
are also presented.}
\begin{ruledtabular}\label{tab:mesons}
\begin{tabular}{ccccccccc}

Mesons ~&~ $B$  ~&~ $B^*$ ~&~  $D$ ~&~  $D^*$  \\
\hline
Exp.$(\mev)$ ~&~ $5279.26\pm0.17$ ~&~  $5325.2\pm0.4$ ~&~ $1864.84\pm0.7$~&~ $2010.26\pm0.07$   \\

This work$(\gev)$~&~ $5.279$~&~  $5.325$ ~&~  $1.863$~&~  $2.010$ \\
\hline
Diquarks~&~ $(bq)_{S=0}$  ~&~ $(bq)_{S=1}$ ~&~  $(cq)_{S=0}$ ~&~  $(cq)_{S=1}$  \\
\hline
This work$(\gev)$~&~ $5.344$~&~  $5.355$ ~&~  $1.963$~&~  $2.00$ \\
QCD sum rules\cite{Kleiv:2013dta}$(\gev)$~&~ $5.08\pm0.04$~&~  $5.08\pm0.04$ ~&~  $1.86\pm0.05$~&~  $1.87\pm0.10$ \\
\end{tabular}
\end{ruledtabular}
\end{table}

%-------------------------------------------------------------------------%
\subsubsection{Color factors in the wave function}
We now turn to discuss the color part of the four-quark states.
The color singlet state of a four-quark system is constructed as following:
\begin{equation}\label{eq:color1}
\left| \bar{3}_{uQ}\otimes3_{\bar{Q}\bar{d}}\right>,  ~~\left| 6_{uQ}\otimes\bar{6}_{\bar{Q}\bar{d}}\right>,
\end{equation}
\begin{equation}\label{eq:color2}
\left|1_{u\bar{d}}\otimes1_{Q\bar{Q}}\right>,  ~~\left|8_{u\bar{d}}\otimes 8_{Q\bar{Q}}\right>,
\end{equation}
\begin{equation}\label{eq:color3}
\left|1_{u\bar{Q}}\otimes1_{Q\bar{d}}\right>, ~~\left|8_{u\bar{Q}}\otimes 8_{Q\bar{d}}\right>,
\end{equation}
which stand as three orthonormal basis-vectors. The expression in Eq.(\ref{eq:color1}) is the so-called tetraquark state with a diquark-anti-diquark structure, we only consider the state $\left| \bar{3}_{uQ}\otimes3_{\bar{Q}\bar{d}}\right>$(denoted as $\left|C_{\text{(tetra)}}\right>$) here\cite{Brodsky:2014xia}, whereas Eq.(\ref{eq:color2}) and Eq.(\ref{eq:color3}) are for the molecular states with a meson-meson structure, specially, the state $\left|1_{u\bar{Q}}\otimes1_{Q\bar{d}}\right>$ (denoted as $\left|C_{\text{(mol)}}\right>$) corresponds to the $B^{(*)}B^{(*)}$ (or $D^{(*)}D^{(*)}$) which is the concern of this work.

The three basis vectors are related to each other through rearrangements \cite{Weinstein:1990gu}
\begin{equation}\label{crelation-1}
\left| 1_{u\bar{Q}}\otimes1_{Q\bar{d}}\right>=\sqrt{\frac{1}{3}}\left|\bar{3}_{uQ}\otimes 3_{\bar{Q}\bar{d}}\right>+\sqrt{\frac{2}{3}}\left|6_{uQ}\otimes\bar{6}_{\bar{Q}\bar{d}}\right>,
\end{equation}
\begin{equation}\label{crelation-2}
\left| 8_{u\bar{Q}}\otimes8_{Q\bar{d}}\right>=-\sqrt{\frac{2}{3}}\left|\bar{3}_{uQ}\otimes 3_{\bar{Q}\bar{d}}\right>+\sqrt{\frac{1}{3}}\left|6_{uQ}\otimes\bar{6}_{\bar{Q}\bar{d}}\right>,
\end{equation}
and
\begin{equation}\label{crelation-3}
\left| 1_{u\bar{d}}\otimes1_{Q\bar{Q}}\right>=-\sqrt{\frac{1}{3}}\left|\bar{3}_{uQ}\otimes 3_{\bar{Q}\bar{d}}\right>+\sqrt{\frac{2}{3}}\left|6_{uQ}\otimes\bar{6}_{\bar{Q}\bar{d}}\right>,
\end{equation}
\begin{equation}\label{crelation-4}
\left| 8_{u\bar{d}}\otimes8_{Q\bar{Q}}\right>=\sqrt{\frac{2}{3}}\left|\bar{3}_{uQ}\otimes 3_{\bar{Q}\bar{d}}\right>+\sqrt{\frac{1}{3}}\left|6_{uQ}\otimes\bar{6}_{\bar{Q}\bar{d}}\right>.
\end{equation}

The color matrix elements which we need in Sec. \ref{sec:Numerical} have been summarized in Table. \ref{tab:Color-elemen}.

\begin{table}[!htbp]
\renewcommand{\arraystretch}{1.4}
\caption{Color matrix elements\cite{symmetry-jVijande}.}
\begin{ruledtabular}\label{tab:Color-elemen}
\begin{tabular}{ccccccc}
 $\hat{O}$ & $(\vec{\lambda}_u\cdot\vec{\lambda}_Q)$ &   $(\vec{\lambda}_{\bar{Q}}\cdot\vec{\lambda}_{\bar{d}})$ & $(\vec{\lambda}_u\cdot\vec{\lambda}_{\bar{Q}})$ & $(\vec{\lambda}_Q\cdot\vec{\lambda}_{\bar{d}})$ & $(\vec{\lambda}_u\cdot\vec{\lambda}_{\bar{d}})$ & $(\vec{\lambda}_Q\cdot\vec{\lambda}_{\bar{Q}})$ \\
\hline
$<\bar{3}_{uQ}3_{\bar{Q}\bar{d}}|\hat{O}|\bar{3}_{uQ}3_{\bar{Q}\bar{d}}>$ & $-8/3$ & $-8/3$ & $-4/3$ & $-4/3$ & $-4/3$ & $-4/3$  \\

$<6_{uQ}\bar{6}_{\bar{Q}\bar{d}}|\hat{O}|6_{uQ}\bar{6}_{\bar{Q}\bar{d}}>$ & $4/3$ & $4/3$ & $-10/3$ & $-10/3$ & $-10/3$ & $-10/3$  \\

$<\bar{3}_{uQ}3_{\bar{Q}\bar{d}}|\hat{O}|6_{uQ}\bar{6}_{\bar{Q}\bar{d}}>$ & $0$ & $0$ & $-2\sqrt{2}$ & $-2\sqrt{2}$ & $2\sqrt{2}$ & $2\sqrt{2}$ \\
\end{tabular}
\end{ruledtabular}
\end{table}

%\bigskip
%-------------------------------------------------------------------------%
\subsubsection{Spin and flavor parts of the wave function}
The flavor and spin parts of molecular states and tetraquark states associated with physical mesons $Z_b^+(10610), \; Z_b^+(10650),\; Z_c^+(3900),\; Z_c^+(4020)$ are listed in Table.~\ref{Tab:spin-flavor},
and the quantum numbers $I^G(J^P)=1^+(1^+)$ for all states.
Specifically,  for the quantum number $(I, I_3)=(1, +1)$ of light flavors $u$ and $\bar d$, the isospin states are $\left|I_\alpha\right>=-u\bar{d}$.
A special note is that our discussion  in introduction and numerical computation made in next section confirm that neither molecular state nor tetraquark correspond to the
observed physical states $Z_b$ and $Z_c$, but their mixtures. Therefore, here using the subscripts [10610], [10650], [3900] and [4020], we only mean that their quantum numbers
correspond to the concerned exotic mesons.

\begin{table}[htbp!]
\renewcommand\tabcolsep{0.19cm}
\renewcommand{\arraystretch}{1.5}

\caption{The flavor  and spin parts of the wave functions for molecular states and tetraquark states, the subscript [10610], [10650], [3900] and [4020] denote that
those pure tetraquark states might be associated with $Z_b^+(10610),Z_b^+(10650),Z_c^+(3900)$ and $Z_c^+(4020)$ respectively.}
\begin{ruledtabular}\label{Tab:spin-flavor}
\begin{tabular}{cccc}
State& Flavor configuration & Spin wave function \\\hline\midrule[0.3pt]
\multirow{1}*{Molecular}
 &$\frac{1}{\sqrt{2}}(B^+\bar{B}^*-B^{*+}\bar{B})$&\multicolumn{1}{c}{$\frac{1}{\sqrt{2}}
\left(0_{b\bar{b}}\otimes1_{u\bar{d}}+1_{b\bar{b}}\otimes0_{u\bar{d}}\right)$ \cite{prd.84-054010} }\\

\multirow{1}*{Tetraquark}
&$(bu)(\bar{b}\bar{d})_{[10610]}$&\multicolumn{1}{c}{$\frac{1}{\sqrt{2}}
\left(0_{bu}\otimes1_{\bar{b}\bar{d}}-1_{bu}\otimes0_{\bar{b}\bar{d}}\right)$ \cite{Ahmed2012prd}}
\\
\hline\midrule[0.3pt]
\multirow{1}*{Molecular}
&$B^{*+}\bar{B}^*$&\multicolumn{1}{c}{$\frac{1}{\sqrt{2}}
\left(0_{b\bar{b}}\otimes1_{u\bar{d}}-1_{b\bar{b}}\otimes0_{u\bar{d}}\right)$ \cite{prd.84-054010}}\\
\multirow{1}*{Tetraquark}
&$(bu)(\bar{b}\bar{d})_{[10650]}$&\multicolumn{1}{c}{$
1_{bu}\otimes1_{\bar{b}\bar{d}}$ \cite{Ahmed2012prd}}
\\
\hline\midrule[0.3pt]
\multirow{1}*{Molecular}
&$\frac{1}{\sqrt{2}}(\bar{D}^*D^++D^{*+}\bar{D^0})$ \cite{Ke2}
&\multicolumn{1}{c}{$
1_{c\bar{c}}\otimes1_{u\bar{d}}$ \cite{12103170}}\\
\multirow{1}*{Tetraquark}
&$(cu)(\bar{c}\bar{d})_{[3900]}$&\multicolumn{1}{c}{$\frac{1}{\sqrt{2}}
\left(0_{cu}\otimes1_{\bar{c}\bar{d}}-1_{cu}\otimes0_{\bar{c}\bar{d}}\right)$ \cite{0412098}}\\

\hline \midrule[0.3pt]
\multirow{1}*{Molecular}
&$D^{*+}\bar{D^*}$ \cite{liu-epjc}
&\multicolumn{1}{c}{$\frac{1}{\sqrt{2}}
\left(0_{c\bar{c}}\otimes1_{u\bar{d}}-1_{c\bar{c}}\otimes0_{u\bar{d}}\right)$}
\\
\multirow{1}*{Tetraquark}&$(cu)(\bar{c}\bar{d})_{[4020]}$
&\multicolumn{1}{c}{$
1_{cu}\otimes1_{\bar{c}\bar{d}}$}\\

\end{tabular}
\end{ruledtabular}
\end{table}

%%%%%%%%%%%%%%%%%%%%%%%%%%%%%%%%%%%%%%%%%%%%%%%%%%%%%%%%%%%%%%%%%%%%%%%%%%%%%%%

\section{Numerical results}
\label{sec:Numerical}

By the Born-Oppenheimer approximation, the binding energy between the two groups (two molecules or diquark-antidiquark) can be written as a function
of distance between $b(c)$ and $\bar b(\bar{c})$  and interactions between the two groups are taken as a perturbation.
Using the wave function described above, we can calculate the binding energy between the two groups with the Heitler-London method.
The binding energy is
\begin{equation}\label{eq:W}
W_\alpha=\left<\psi_{\alpha}\left|H^{\alpha}_{\text{int}}\right|\psi_{\alpha}\right>,%~~~~\alpha=1, 2,
\end{equation}
where $H^\alpha_{\text{int}}$ (see Sec.~\ref{potentials} for details) is a perturbative term for the molecular structure and as well as for the tetraquark.

\label{results}
In this work, we take the meson-quark coupling constants $g_{ch}$ and cut-off parameters $\Lambda_\chi$ from Ref. \cite{Vijande:2004he}, and the masses of light mesons are taken  the
databook (PDG) values \cite{Agashe:2014kda}, then the other parameters, like $b, c, h, \alpha_0$ etc, have been determined by
fitting the heavy mesons spectra (see Table. \ref{tab:mesons}).  They are presented in Table. \ref{tab:parameters}.

\begin{table}[!htbp]
\renewcommand{\arraystretch}{1.3}
\caption{The parameters of the model and masses of concerned mesons.}
\begin{ruledtabular}\label{tab:parameters}
\begin{tabular}{ccccccc}

$m_{u( d)}$ ~&~ $m_b$ ~&~ $m_c$ ~&~ $\mu_\sigma$ ~&~ $\mu_{\pi}$ ~&~ $\mu_{\eta}$ \\

$0.313\gev$ ~&~ $4.80\gev$ ~&~ $1.40\gev$ ~&~ $490\mev$ ~&~ $139.57\mev$ ~&~ $547.862\mev$\\
\hline
$\mu_\rho$ ~&~$g_{ch}^2/4\pi$  ~&~ $\Lambda_\pi$  ~&~ $\Lambda_\sigma$ ~&~ $\Lambda_\eta$ ~&~ $\Lambda_0$  \\

$775.26\mev$ ~&~ $0.54$~&~ $4.2fm^{-1}$  ~&~ $4.2fm^{-1}$ ~&~ $5.2fm^{-1}$ ~&~ $0.113 fm^{-1}$\\
\hline
$\mu_0$ ~&~ $\alpha_0$~&~$h$ ~&~b ~&~$c$ ~&~  $$   \\

$36.976\mev$ ~&~ $2.118$~&~ $0.79\gev$ ~&~ $0.148\gev^2$~&~  $-0.319\gev$ ~&~  $$ \\

\end{tabular}
\end{ruledtabular}
\end{table}
\bigskip

\subsection{Molecular structure}

\begin{figure}[!htbp]
\begin{center}
\begin{tabular}{ccc}
\scalebox{0.305}{\includegraphics{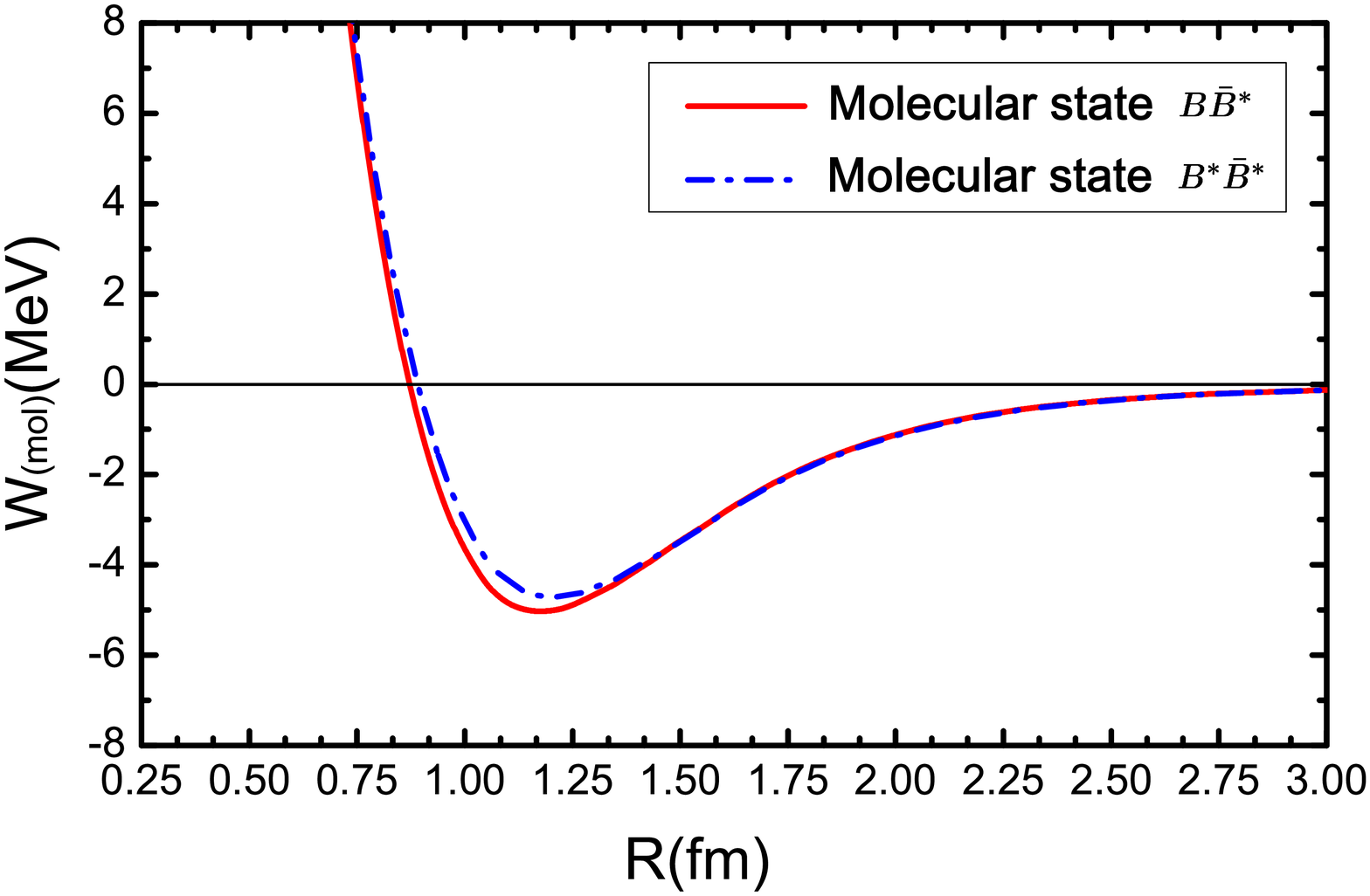}}\scalebox{0.305}{\includegraphics{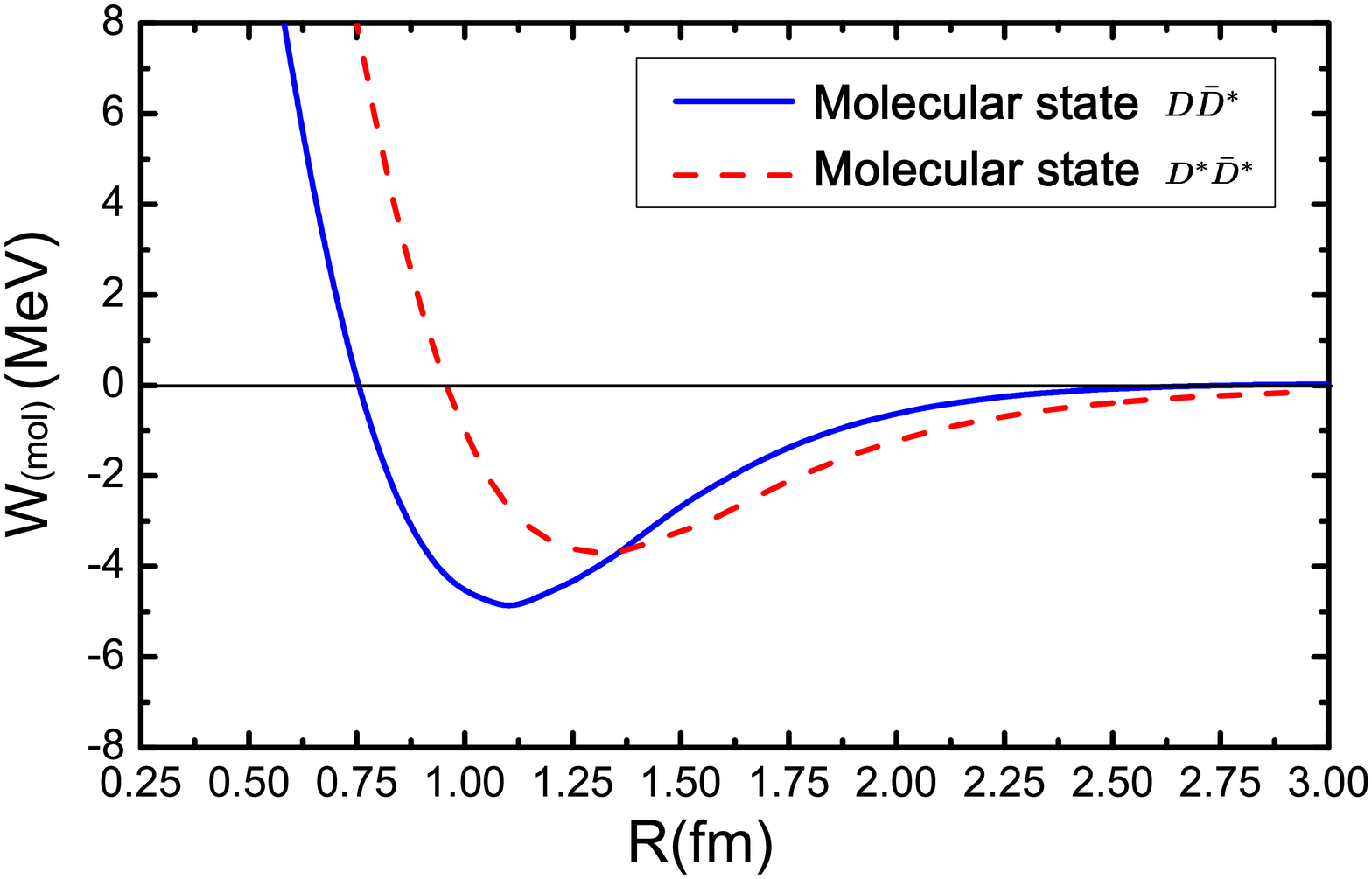}}
\\
(I)\qquad\qquad\qquad\qquad\qquad\qquad\qquad\qquad\qquad\qquad
\,\,\,\,\,\,\,\,\,\,\,\,\,\,\,\,\,\,\,\,\,\,\,\,\,\,(II)
\end{tabular}
\end{center}
\caption{The obtained binding energy for $B\bar B^*, B^*\bar B^*, D\bar D^*$ and $D^*\bar D^*$  molecular structures.}\label{Fig1}
\end{figure}

In this subsection, we discuss the case of the molecular structure.
In terms of the obtained wave functions and eigen-energies  of the two constituent mesons, we estimate the expectation values shown in Eq.(\ref{eq:W}).
The color, spin and flavor parts of Eq.(\ref{eq:W}) are shown  in Table. \ref{tab:Color-elemen} and  Table. \ref{Tab:spin-flavor}, and the integration of the radial part is carried out numerically. The binding energy of molecular states $W_{(\text{mol})}$ versus the distance between $Q$ and $\bar Q$ is
drawn in Fig. \ref{Fig1}. The figures indicate that there exist minima $E_{\text{(mol)}}$ for all the concerned states. As we expect, in the Born-Oppenheimer approximation, a molecular
state of the four-quark system possesses a minimum which corresponds to a stable structure.
Then, the masses of molecular states $M_{\text{(mol)}}=m_1+m_2+E_{\text{(mol)}}$ , where $m_1$ and $m_2$ are the masses of the constituent mesons, are presented in Table.\ref{tab:minima}.

Here we define $R$ as the distance between $Q$ and $\bar Q$. The minima are located at around $R\sim 1 $fm, and the $B^{(*)}$-$\bar B^{(*)}, \; D^{(*)}$-$\bar D^{(*)}$ structures can be considered as loosely bound states with binding energies of $-3\sim-5\mev$.

\begin{table}[!htbp]
\renewcommand{\arraystretch}{1.45}
\caption{Binding energy minima($E_{\text{(mol)}} (\mev)$), distance $R (fm)$ between $Q\bar{Q}$ and the calculated masses $M_{\text{(mol)}} (\mev)$ of $B\bar B^*, B^*\bar B^*, D\bar D^*$ and $D^*\bar D^*$  molecular structures.}
\begin{ruledtabular}\label{tab:minima}
\begin{tabular}{ccc|ccc|ccc|cccc}
\multicolumn{3}{c}{$(\bar bu)(b\bar d)_{B\bar B^*}$}&\multicolumn{3}{c}{$(\bar bu)(b\bar d)_{B^*\bar B^*}$}&
\multicolumn{3}{c}{$(\bar cu)(c\bar d)_{D\bar D^*}$}&\multicolumn{3}{c}{$(\bar cu)(c\bar d)_{D^*\bar D^*}$}\\
\hline
$R$& $E_{\text{(mol)}}$ &$M_{\text{(mol)}}$ & $R$ & $E_{\text{(mol)}}$&$M_{\text{(mol)}}$ & $R$ & $E_{\text{(mol)}}$&$M_{\text{(mol)}}$
& $R$ & $E_{\text{(mol)}}$&$M_{\text{(mol)}}$\\
\hline
$1.17$ & $-5.034$ & $10598.966$ &$1.2$ & $-4.717$ & $10645.283$ &$1.15$  & $-4.909$ & $3868.091$   &$1.35$ & $-3.705$  & $4016.295$  \\
\end{tabular}
\end{ruledtabular}
\end{table}

\subsection{The Tetraquark structure}
\begin{figure}[!htbp]
\begin{center}
\begin{tabular}{ccc}
\scalebox{0.305}{\includegraphics{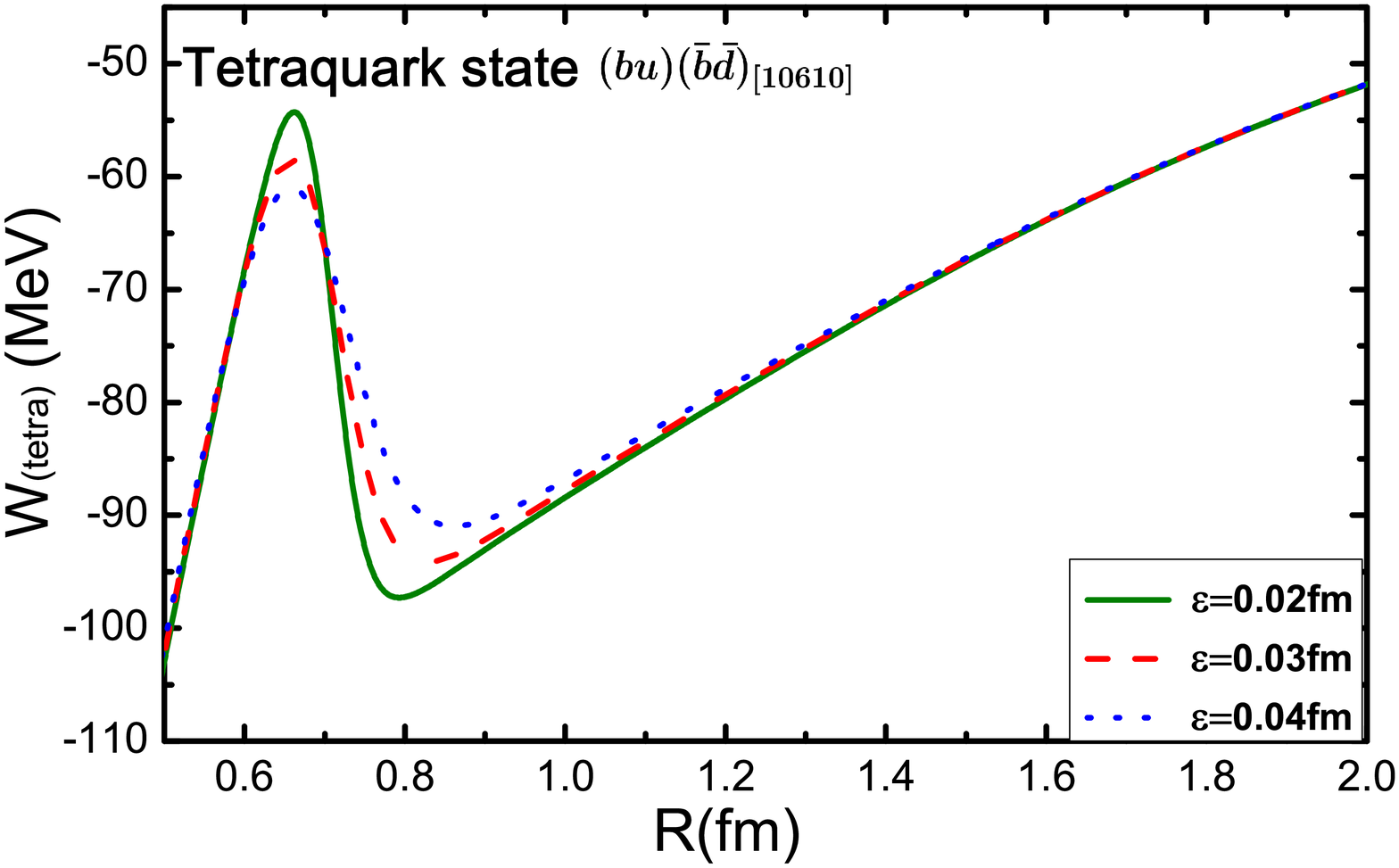}}\scalebox{0.305}{\includegraphics{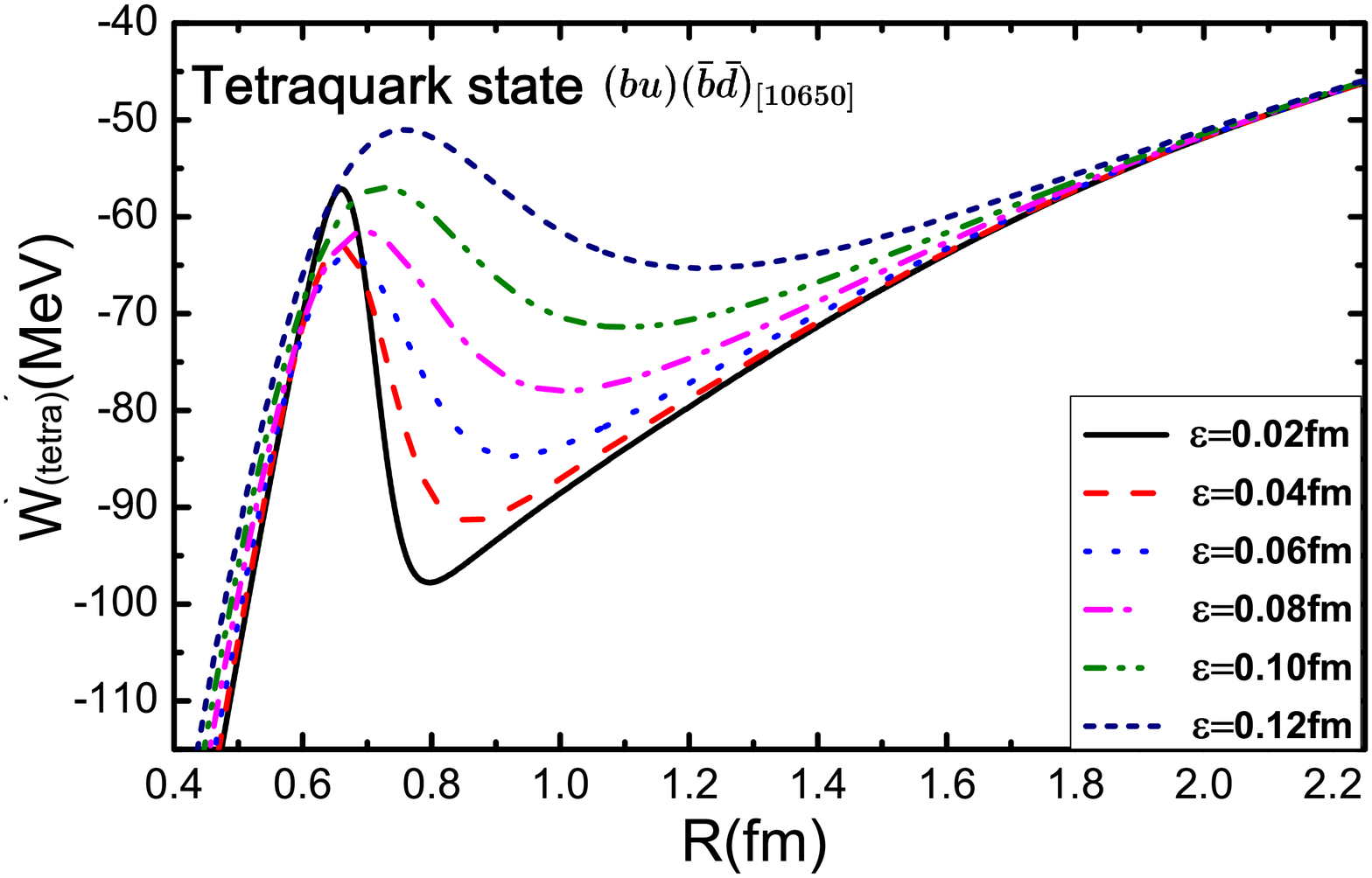}}
\\
(I)\qquad\qquad\qquad\qquad\qquad\qquad\qquad\qquad\qquad\qquad
\,\,\,\,\,\,\,\,\,\,\,\,\,\,\,\,\,\,\,\,\,\,\,\,\,\,(II)
\end{tabular}
\end{center}
\caption{ The variation of the obtained binding energy for $(bu)(\bar{b}\bar{d})_{[10610]},\;(bu)(\bar{b}\bar{d})_{[10650]}$ tetraquark structures in $\varepsilon$.
Here, the values of $\varepsilon$ are taken as 0.02 to 0.12fm.}\label{Fig3}
\end{figure}

\begin{figure}[!htbp]
\begin{center}
\begin{tabular}{ccc}
\scalebox{0.305}{\includegraphics{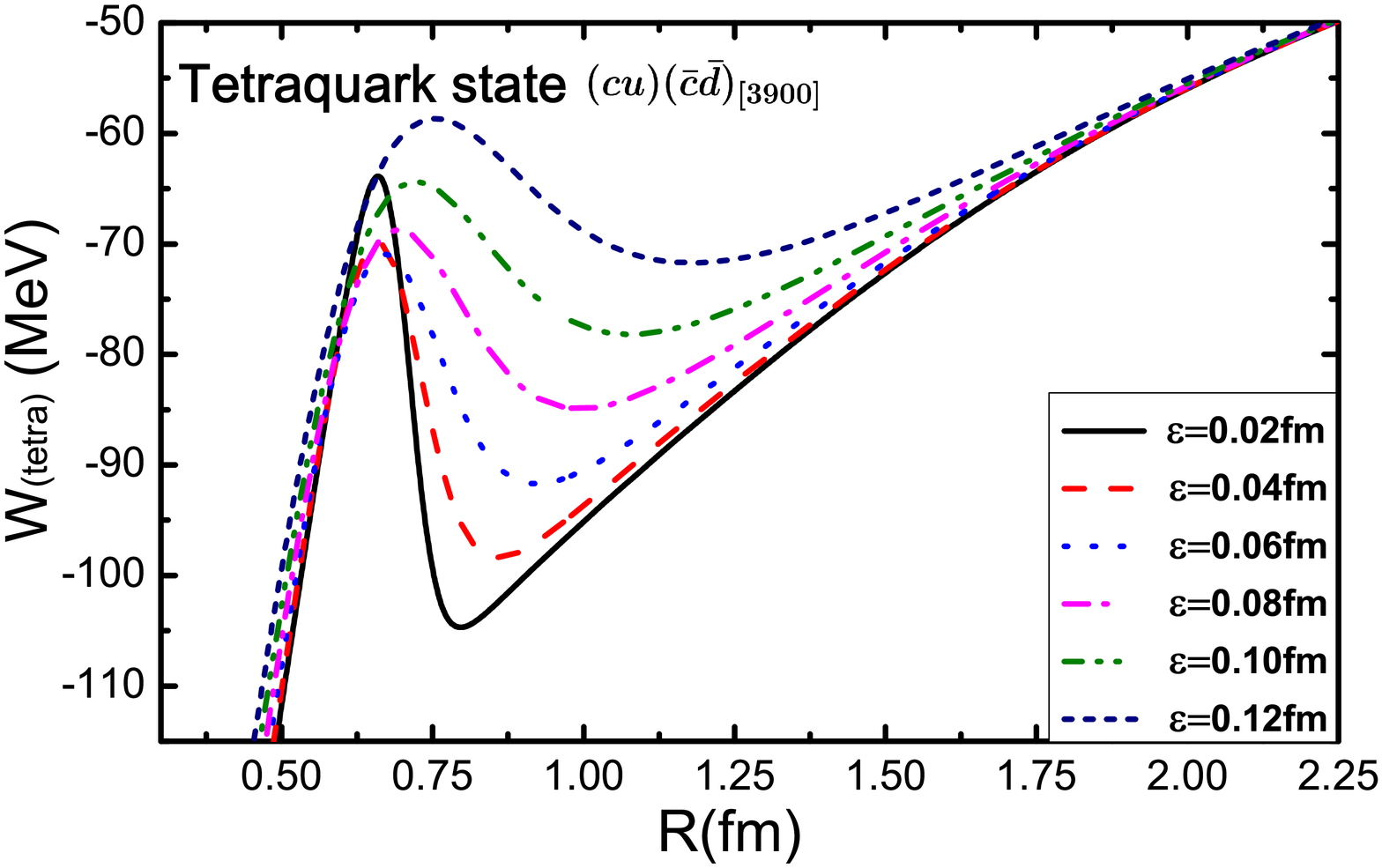}}\scalebox{0.305}{\includegraphics{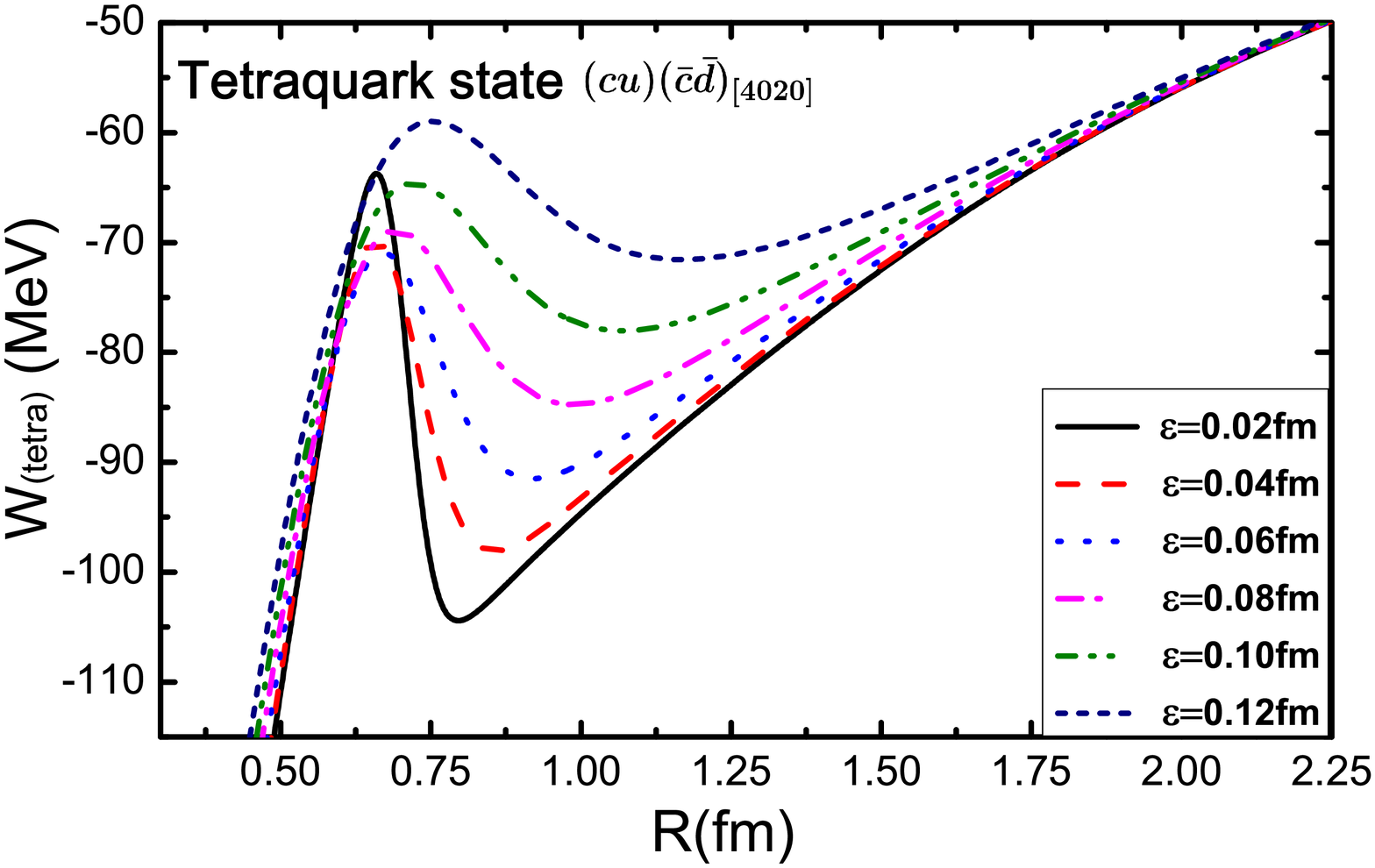}}
\\
(III)\qquad\qquad\qquad\qquad\qquad\qquad\qquad\qquad\qquad\qquad
\,\,\,\,\,\,\,\,\,\,\,\,\,\,\,\,\,\,\,\,\,\,\,\,\,\,(IV)
\end{tabular}
\end{center}
\caption{The obtained binding energies for $(cu)(\bar{c}\bar{d})_{[3900]}$ and $(cu)(\bar{c}\bar{d})_{[4020]}$ tetraquark structures with $\varepsilon$.
Here, the values of $\varepsilon$ are taken as  0.02 to 0.12fm.}\label{Fig4}
\end{figure}

Now let us turn to discuss the case of tetraquark.
With the same procedure as for the molecular states, we obtain dependence of the binding energies of tetraquark  $W_{\text{(tetra)}}$  on the distance between $Q$ and $\bar{Q}$ with various values of the parameter $\varepsilon$. The results are shown in Fig. \ref{Fig3} and Fig. \ref{Fig4}.

It is interesting, we find that there indeed exists minimum $E_{\text{(tetra)}}$ with respect to the distance between $Q$ and $\bar Q$, and the stable point
corresponds to the distance at $R\approx 0.79\sim 1.5$ fm which is comparable with that for molecular states, but generally shorter. It seems reasonable.
The masses of the tetraquark (defined as $M_{\text{(tetra)}}=m_{D1}+m_{D2}+E_{\text{(tetra)}}$), where $m_{D1}$ and
$m_{D2}$ are the masses of the diquark and anti-diquark, are presented in Table.~\ref{tab:Tetra-spectra}.

\begin{table}[!htbp]
\renewcommand{\arraystretch}{1.35}
\caption{Binding energy minima($E_{\text{(tetra)}} (\mev)$), distance $R (fm)$ between $Q\bar{Q}$ and the calculated masses $M_{\text{(tetra)}} (\mev)$ of $(bu)(\bar{b}\bar{d})_{[10610]},\;(bu)(\bar{b}\bar{d})_{[10650]},\; (cu)(\bar{c}\bar{d})_{[3900]}$ and $(cu)(\bar{c}\bar{d})_{[4020]}$ tetraquark structures, with
respect to the free parameter $\varepsilon(fm)$ .}
\begin{ruledtabular}\label{tab:Tetra-spectra}
\begin{tabular}{c|ccc|ccc|ccc|cccc}
\multicolumn{4}{c}{~~~~~~~$(bu)(\bar b \bar d)_{[10610]}$}&\multicolumn{3}{c}{$(bu)(\bar b \bar d)_{[10650]}$}&
\multicolumn{3}{c}{$(cu)(\bar c \bar d)_{[3900]}$}&\multicolumn{3}{c}{$(cu)(\bar c \bar d)_{[4020]}$}\\
\hline
$\varepsilon$ &$R$& $E_{\text{(tetra)}}$ &$M_{\text{(tetra)}}$ & $R$ & $E_{\text{(tetra)}}$&$M_{\text{(tetra)}}$ & $R$ & $E_{\text{(tetra)}}$&$M_{\text{(tetra)}}$& $R$ & $E_{\text{(tetra)}}$&$M_{\text{(tetra)}}$\\
\hline
$0.02$ &$0.79$ &$-97.387$ &$10601.613$ &$0.79$ &$-98.158$ &$10611.842$ &$0.79$  &$-105.080$ & $3857.92$   &$0.79$ & $-104.813$  & $3895.187$  \\
$0.03$ &$0.82$ &$-94.181$ &$10604.819$ &$0.82$ &$-94.846$ &$10615.154$ &$0.82$  &$-101.780$ & $3861.22$   &$0.82$ & $101.533$  & $3898.467$  \\
$0.04$ &$0.865$ & $-90.99$ & $10608.010$ &$0.85$ & $-91.510$ & $10618.490$ &$0.85$  & $-98.495$ & $3864.505$  &$0.85$ & $-98.241$  & $3901.759$  \\
$0.05$    &$$  & $$ & $$ &$0.88$ & $-88.136$ & $10621.864 $ &$0.88$ & $-95.146$ & $3867.854$  &$0.88$ & $-94.916$  & $3905.084$  \\
$0.06$ &$$  & $$ & $$ &$0.94$ & $-84.735$ & $10625.265$ &$0.91$ & $-91.753$ & $3871.247$  &$0.91$ & $-91.552$  & $3908.448$  \\
$0.07$    &$$  & $$ & $$ &$0.97$ & $-81.380$ & $10628.62 $ &$0.97$ & $-88.335$ & $3874.665$   &$0.94$ & $-88.157$  & $3911.843$  \\
$0.08$ &$$  & $$ & $$ &$1.0$ & $-78.010$ & $10631.990$ &$1.0$ & $-84.925$ & $3878.048$  &$1.0$ & $-84.767$  & $3915.233$  \\
$0.09$    &$$  & $$ & $$ &$1.06$ & $-74.661$ & $10635.339 $ &$1.03$ & $-81.562$ & $3881.438$  &$1.03$ & $-81.408$  & $3918.592$  \\
$0.10$   &$$  & $$ & $$ &$1.09$ & $-71.421$ & $10638.579 $ &$1.09$ & $-78.180$ & $3884.820$   &$1.06$ & $-78.052$  & $3921.948$  \\
$0.11$ &$$  & $$ & $$ &$1.15$ & $-68.289$ & $10641.711$ &$1.12$ & $-74.911$ & $3888.089$  &$1.12$ & $-74.760$  & $3925.234$  \\
$0.12$ &$$  & $$ & $$ &$1.21$ & $-65.296$ & $10644.704$ &$1.18$ & $-71.717$ & $3891.283$  &$1.18$ & $-71.555$  & $3928.445$  \\
$0.13$ &$$  & $$ & $$ &$1.30$ & $-62.463$ & $10647.537$ &$1.24$ & $-68.646$ & $3894.354$  &$1.24$ & $-68.468$  & $3931.532$  \\
$0.14$ &$$  & $$ & $$ &$1.36$ & $-59.826$ & $10650.174$ &$1.3$ & $-65.720$ & $3897.280$  &$1.30$ & $-65.530$  & $3934.470$  \\
$0.15$ &$$  & $$ & $$ &$$ & $$ & $$ &$$ & $$ & $$  &$1.36$ & $-62.748$  & $3937.252$  \\
$0.16$ &$$  & $$ & $$ &$$ & $$ & $ $ &$$ & $$ & $$  &$1.45$ & $-60.138$  & $3939.862$  \\
$0.17$ &$$  & $$ & $$ &$$ & $$ & $ $ &$$ & $$ & $$  &$1.51$ & $-57.705$  & $3942.295$  \\
\end{tabular}
\end{ruledtabular}
\end{table}

%%%%%%%%%%%%%%%%%%%%%%%%%%%%%%%%%%%%%%%%%%%%%%%%%%%%%%%%%%%%%%%%%%%%%%%%%%%%%%%
\section{Conclusion and discussion}
\label{sec:conclusion}

As discussed in the introduction, many authors suggested that the newly observed four-quark states $Z_b(10610),\; Z_b(10650),\;Z_c(3900),\;Z_c(4020)$ etc. are hadronic
molecules, the reason is that their masses are close to the sum of some mesons $B,\;B^*,\; D,\; D^*$. However, for all of them the sum of the masses of the constituent mesons
is smaller than the  mass of the concerned exotic meson. By the potential model, the binding energy should be negative, and the calculated values of the binding energies
shown in Table. \ref{tab:minima} confirm the allegation. Therefore, assuming them to
be molecular states bring up an inconsistency. To solve this puzzle, there must be corresponding tetraquark states which mix with the molecular states to result in
the observed physical hadrons.

The possible energy matrix is written as
\begin{equation}
H=\left(\begin{array}{cc}
M_{\text{(mol)}} & \Delta_Q \\
\Delta_Q & M_{\text{(tetra)}}
\end{array}\right),
\end{equation}
where $M_{\text{(mol)}}$ and $M_{\text{(tetra)}}$ are the masses of a pure molecular state and a tetraquark calculated in the theoretical framework, the off-diagonal element $\Delta_Q$
whose subscript $Q$ means that
it may be flavor-dependent ($b$ or $c$), is a mixing parameter. % which so far we do not know how to evaluate yet.
Solving the secular equation:
\begin{equation}
\left| \begin{array}{cc}
M_{\text{(mol)}}-\lambda & \Delta_Q \\
\Delta_Q & M_{\text{(tetra)}}-\lambda
\end{array}\right|=0
\end{equation}
we obtain two eigenvalues
\begin{equation}\label{eigenvalue}
\lambda_{\pm}={M_{\text{(mol)}}+M_{\text{(tetra)}}\pm\sqrt{(M_{\text{(mol)}}-M_{\text{(tetra)}})^2+4\Delta_Q^2}\over 2},
\end{equation}
and $\lambda_{\pm}$ are the masses of physical states i.e. mixtures of molecular states and tetraquarks.

It is noted that $\lambda_{+}$$>\textit{Max}(M_{\text{(mol)}}, M_{\text{(tetra)}})$ and $\lambda_{-}$$< \textit{Min}(M_{\text{(mol)}}, M_{\text{(tetra)}})$.
In our framework, the masses of both molecular states and tetraquark states are below that of the observed exotic mesons, so
we expect that $\lambda_+$'s correspond to the physical exotic states which are the experimentally observed $Z_b(10610),\;Z_b(10650)$, $Z_c(3900)$  and $Z_c(4020)$.
If so, it is natural to predict existence of their partner exotic states whose masses are $\lambda_-$'s smaller than the observed states as
listed in Table.\ref{tab:Predicted-spectra}.
\begin{table}[!htbp]
\renewcommand{\arraystretch}{1.35}
\caption{The mixing parameter $\Delta_Q$ and the masses $M'$ of the predicted counterparts of  $Z_b(10610), Z_b(10650), Z_c(3900)$ and $Z_c(4020)$, with
respect to the free parameter $\varepsilon$ .}
\begin{ruledtabular}\label{tab:Predicted-spectra}
\begin{tabular}{c|cc|cc|cc|cc}
\multicolumn{3}{c}{~~~~~~~$Z_b(10610)$}&\multicolumn{2}{c}{$Z_b(10650)$}&
\multicolumn{2}{c}{$Z_c(3900)$} &\multicolumn{2}{c}{$Z_c(4020)$}   \\
\hline
$\varepsilon(fm)$  &$\Delta_b(\mev)$ &$M'(\mev)$& $\Delta_b(\mev)$ &$M'(\mev)$ &$\Delta_c(\mev)$ &$M'(\mev)$ &$\Delta_c(\mev)$ &$M'(\mev)$ \\
\hline
$0.02$  &$8.0$  & $10592.2$ &$18.10$ & $10603.9$ &$35.63$& $3827.01$ &$29.04$& $3888.58$  \\
$0.03$  &$5.81$ & $10595.4$ &$17.36$ & $10607.2$ &$34.17$& $3830.31$ &$28.67$& $3891.86$  \\
$0.04$  &$1.92$ & $10598.6$ &$16.58$ & $10610.6$ &$32.65$& $3833.60$ &$28.29$& $3895.15$  \\
$0.05$  &$$     & $$        &$15.75$ & $10613.9$ &$31.03$& $3836.94$ &$27.90$& $3898.48$  \\
$0.06$  &$$     & $$        &$14.87$ & $10617.3$ &$29.29$& $3840.34$ &$27.49$& $3901.84$  \\
$0.07$  &$$     & $$        &$13.95$ & $10620.7$ &$27.43$& $3843.76$ &$27.08$& $3905.24$  \\
$0.08$  &$$     & $$        &$12.96$ & $10624.1$ &$25.45$& $3847.14$ &$26.67$& $3908.63$  \\
$0.09$  &$$     & $$        &$11.89$ & $10627.4$ &$23.30$& $3850.53$ &$26.25$& $3911.99$  \\
$0.10$  &$$     & $$        &$10.76$ & $10630.7$ &$20.94$& $3853.91$ &$25.82$& $3915.34$  \\
$0.11$  &$$     & $$        &$9.54$  & $10633.8$ &$18.36$& $3857.18$ &$25.40$& $3918.63$  \\
$0.12$  &$$     & $$        &$8.20$  & $10636.8$ &$15.44$& $3860.37$ &$24.98$& $3921.84$  \\
$0.13$  &$$     & $$        &$6.70$  & $10639.6$ &$11.98$& $3863.44$ &$24.57$& $3924.93$  \\
$0.14$  &$$     & $$        &$4.89$  & $10642.3$ &$7.29$ & $3866.37$ &$24.17$& $3927.87$  \\
$0.15$  &$$     & $$        &$$    & $ $         &$$    & $$         &$23.78$& $3930.65$  \\
$0.16$  &$$     & $$        &$$    & $ $         &$$    & $$         &$23.42$& $3933.26$  \\
$0.17$  &$$     & $$        &$$    & $ $         &$$    & $$         &$23.07$& $3935.69$  \\
\end{tabular}
\end{ruledtabular}
\end{table}

In this scheme, we conclude that the tetraquark states must exist.

Our numerical results indicate that for both molecule and tetraquark states, the functions of the binding energies possess minima. For the case of molecular states, the
minimum occurs at $R\sim1$ fm (for $Z_b$ and $Z_c$, see Table. \ref{tab:minima}), whereas, for the tetraquark, $R$=0.79$ \sim$1.5 fm depending on the parameter $\varepsilon$
where $R$ is the distance between $Q$ and $\bar Q$.
The situations for $Z_b$ and $Z_c$ are slightly different, but roughly the tendency is the same. It is also noted that the resultant  $R$ is flavor dependent, but no matter for $c$ or $b$, it falls within a reasonable range i.e. roughly $1/\Lambda_{QCD}$.
\begin{figure}[!htbp]
\begin{center}
\begin{tabular}{ccc}
\scalebox{1.25}{\includegraphics{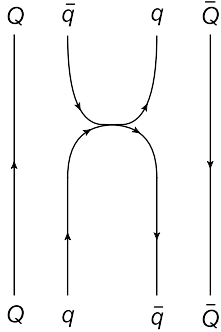}}
\\
\end{tabular}
\end{center}
\caption{Mixing Mechanism}\label{fig:mix}
\end{figure}

A few observations on the results. First from Fig. \ref{Fig3} and Fig. \ref{Fig4}, one notices that the local minimum is a metastable one and for $R<0.6$ fm, the binding energy drops drastically.
It may imply that there could be an anarchy state for a four-quark system. But it is only a qualitative inference, then the computed value for the binding energy
is not reliable because here the adopted picture is only valid for the diqark-anti-diquark structure instead of the anarchy state.

The main conclusion is that there are minima for both molecule and tetraquark structures, so that both of them can exist, a natural mixture would be expected.
The mixing between molecular structure and tetraquark is induced by exchanging quark and anti-quark which reside in different groups (Fig. \ref{fig:mix}). Such mechanism has been
discussed in literature\cite{He:2005eya}, but because it is a non-perturbative QCD effect, there is no appropriate way to
calculate $\Delta_Q$ yet. However,  we may fix  it phenomenologically, for example, using the
values given in Table. \ref{tab:minima}, Table. \ref{tab:Tetra-spectra} and $\lambda_+$ of Eq.(\ref{eigenvalue}), we obtain $\Delta_Q=2\sim 35 ~\mev$.

With the provided picture, we predict the positions of the partners of $Z_b(10510)$, $Z_b(10650)$, $Z_c(3900)$ and $Z_c(4020)
(Z_c(4025))$ which weakly depend on the value of $\varepsilon$. Therefore, the key point to validate or negate our picture is to look for the counter-partners of
the observed exotic mesons. However, since the masses of the expected mesons are below the production thresholds of $B^{(*)}$-$\bar B^{(*)}$ or $D^{(*)}$-$\bar D^{(*)}$
(which can be realized in $Z_b$ decays but not in $Z_c$'s), one should look for them in the decay modes such $KK\pi$ etc.

To require quantitatively reliable conclusion, more information (theoretical and especially experimental) is needed.
Indeed the more accurate data are being accumulated, and  we hope that more measurements will be
carried out at BES, SuperBelle and LHCb, as well as the other proposed colliders.

\section*{ACKNOWLEDGMENTS}

We sincerely thank HY Cheng for helpful discussions.
This work is supported by the National Natural Science Foundation
of China (NNSFC) under the contract No. 11375128.

%%%%%%%%%%%%%%%%%%%%%%%%%%%%%%%%%%%%%%%%%%%%%%%%%%%%%%%%%%%%%%%%%%%%%%%%%%%%%%%%%%%%%%%%%%%%%%%%%
%%%%%%%%%%%%%%%%%%%%%%%%%%%%%%%%%%%%%%%%%%%%%%%%%%%%%%%%%%%%%%%%%%%%%%%%%%%%%%%%%%%%%%%%%%%%%%%%%

\end{document}